\documentclass{aa}
\usepackage[varg]{txfonts}

\usepackage{graphicx}
\usepackage{amssymb,amsmath,pdflscape}
\usepackage{natbib}
\usepackage{empheq}
\usepackage{subeqnarray}
\usepackage{mathtools}
\usepackage{upgreek}
\usepackage{cases}

\begin{document}

\title{Existence of slowly rotating bipolytropes with prolate cores}

\titlerunning{Existence of slowly rotating bipolytropes with prolate cores}

\author{C. Staelen\inst{1} \and J.-M. Hur\'e\inst{1} \and A. Meunier\inst{2} \and P. No\'e\inst{1,3}}
\institute{Univ. Bordeaux, CNRS, LAB, UMR 5804, F-33600 Pessac, France \\ \email{clement.staelen@u-bordeaux.fr} 
\and Univ. Luxembourg, Department of Mathematics, L-4364 Esch-sur-Alzette, Luxembourg
\and Univ. Strasbourg, T\'el\'ecom Physique Strasbourg, F-67400 Illkirch-Graffenstaden, France}

\date{Received ??? / Accepted ???}

\abstract{We report the existence of hydrostatic equilibrium states for a composite body made of two rigidly rotating, homogeneous layers bounded by spheroidal surfaces, where the core has a prolate shape. These new configurations require an oblate envelope that spins faster than the core. No solution with a prolate envelope is found. For some parameters, the prolate core can even be at rest. Numerical experiments based on the self-consistent field method support this result in the case of heterogeneous layers with polytropic equations of state. The possible cancellation of the first gravitational moment, $J_2$, is discussed.}


\maketitle

\section{Introduction}

A common property of self-gravitating systems carrying angular momentum is the natural propensity to flatten, which is indisputably observed for planets, stars, and galaxies. The centrifugal force is actually extremely efficient in pushing matter away from the rotation axis. As calculated by Maclaurin for a rigidly rotating, homogeneous body, the flattening takes the form of a spheroid (i.e. an ellipsoid of revolution) spinning around its minor axis. Oblateness is also observed for compressible matter \citep{hachisu86}. Prolate configurations corresponding to an elongation along the rotation axis have always been ruled out when only gravity, pressure forces, and rotation are considered. Prolateness can be provoked by additional mechanisms or ingredients, such as meridional circulation \citep{fe14} or magnetic fields \citep[e.g.][]{oa72,lj09,ktye11,hs24}. 

While prolateness is not permitted in Maclaurin's theory, we show in this article that a homogeneous prolate spheroid can be a figure of equilibrium if it is embedded in an envelope externally bounded by an oblate spheroidal surface. This effect, which is demonstrated in the slow-rotating and rigidly rotating limits, is purely hydrodynamical and requires asynchronous motion between the two layers. Based upon numerical simulations, we reach a similar conclusion when the two components are inhomogeneous. This result, which contributes to the theory of figures, provides new tools for studying the formation and evolution of piecewise homogeneous systems like rocky and/or icy planets, planetesimals, and asteroids in the Solar System \citep[e.g.][]{rcc15}. Furthermore, it could also help explain the presence of prolate molecular clouds in the interstellar medium \citep[e.g.][]{fp00,ct10}.

\section{Rotating bodies with spheroid bounding layers}

The hydrostatic equilibrium of a rigidly rotating fluid mass made of several layers bounded by spheroidal surfaces was examined in detail by \citet{hamy89,hamy90}. Assuming the mass density was monotonic with depth, he demonstrated that all the surfaces are necessarily confocal to each other (i.e. all ellipses share the same foci) and all spheroids necessarily rotate asynchronously \citep[see also][]{veronet12,moulton16,mmc83}. These conditions constitute Hamy's `no-go' theorem.

Out of confocality, the interfaces between layers cannot be exact spheroids, but numerical experiments show that the departure is weak for a wide range of configurations. This led \citet{h2022a} to investigate the approximation of spheroidal layers for fast rotators. The nested spheroidal figures of equilibrium (NSFoE) theory reported therein assumes that hydrostatic equilibrium is satisfied at the pole and at the equator of each layer. The NSFoE theory is expected to be reliable when all spheroids are close to confocal with each other. As a consequence, this theory goes beyond the slowly rotating regime usually considered \citep{clairaut43,hamy89} and reliably describes fast rotators. With this approach, any relative motion of layers is possible, from global rotation, in which all the rotation rates ($\varOmega_i$) are the same, to a fully asynchronous regime, in which the $\varOmega_i$'s are all different. 

In all these theories, only oblate configurations have been obtained or reported\footnote{Note that prolate subsystems were already considered by \citet{dalembert47} in the context of a rigid core supporting a shallow ocean.}. Relaxing this assumption opens up to new states. In this work, we return to the slowly rotating limit, which enables us to bypass the complexity of the general formulae and to derive analytical criteria for the existence of prolate subcomponents.

\section{Equation set relevant to a two-layer configuration}
\label{sec:twolayerconf}

\begin{figure}[ht]
       \centering
       \includegraphics[width=0.67\linewidth]{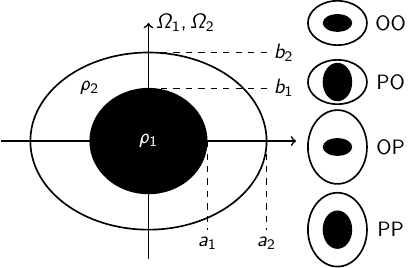}
       \caption{Heterogeneous system made of two homogeneous layers bound by spheroidal surfaces, namely the core (black) and the envelope (white). Four core-envelope configurations are analysed. O stands for oblate and P for prolate; the first letter refers to the core and the second to the envelope.}
       \label{fig:system}
\end{figure}

Our assumptions and notations are the same as those in  \citet{h2022a}. We considered a heterogeneous body made of two homogeneous, spheroidal components, as pictured in Fig.~\ref{fig:system}. Let $\rho_i$, $\varOmega_i$, $a_i$, and $b_i$ be the mass density, the rotation rate, the equatorial radius, and the polar radius of layer $i$, respectively. Index 1 is for the deepest layer (hereafter `the core') and index 2 for the upper layer (hereafter `the envelope'). We define the relative equatorial radius of the core ${q=a_1/a_2}$, the mass-density jump at the interface ${\alpha=\rho_1/\rho_2}$, the squared eccentricity of layer $i$, ${\varepsilon_i^2=1-(b_i/a_i)^2}$, and the dimensionless squared rotation rate of layer $i$,  ${\tilde\varOmega_i^2 = \varOmega_i^2/(2\uppi G\rho_2)}$, where $G$ is the gravitational constant. Cases with $q=0$, $q=1$, or $\alpha=1$ have no special interest, as they all correspond to a single-component body. Asynchronous motion is allowed: $\varOmega_1$ and $\varOmega_2$ are not necessarily equal. As mentioned above, the NSFoE theory is reliable when the spheroids are nearly confocal (i.e. when the confocal parameter, $c=q^2\varepsilon_1^2-\varepsilon_2^2$, is small enough). The error is typically below 1\% for $|c|\lesssim 0.5$.

The starting points of the present analysis are Eqs. (66a) and (66b) from \citet{h2022a}, which give the rotation rates at equilibrium of each layer in the regime of slow rotations, in the case of oblate interfaces. At the first order in $\varepsilon_i^2$, we have
\begin{subequations}\label{eq:om2_sol}
       \begin{empheq}[left={\empheqlbrace\,}]{align}
              &\tilde\varOmega_2^2 = \frac{4}{15}\varepsilon_2^2 - \frac{2}{15}(\alpha-1) q^3\left(3q^2\varepsilon_1^2-5\varepsilon_2^2\right), \label{eq:om22_sol}\\
              &\alpha\tilde\varOmega_1^2 = \tilde\varOmega_2^2 + \frac{2}{15}(\alpha-1)\left[2(\alpha-1)\varepsilon_1^2 + 5\varepsilon_1^2-3\varepsilon_2^2\right]. \label{eq:om12_sol}
       \end{empheq}
\end{subequations}

\noindent We note that these equations are not valid for $q=0$ or $q=1$, as one of the layers does not exist. The equations have been established in the case of two oblate interfaces, but they still hold for prolate interfaces (i.e. $\varepsilon_i^2<0$). This is due to the mathematical continuity in the expression for gravitational potential when the eccentricity becomes an imaginary number.

Thus, we worked with the same equations for a prolate-oblate mixture. There is therefore a maximum of four types of configurations, as shown in Fig.~\ref{fig:system}: an oblate core and an oblate envelope (OO), an oblate core with a prolate envelope (OP), a prolate core and an oblate envelope (PO), and a prolate core with a prolate envelope (PP).

The first case is already treated in detail in \citet{h2022a}, where it is shown that there are two families of solutions. The first is when rotations are synchronous, or $\varOmega_1 = \varOmega_2$. This requires that the core-envelope interface is a surface of constant pressure (type-C solutions). The mass-density jump is not free (see Eq.~\eqref{eq:alphac} below). The second is when rotations are asynchronous, or $\varOmega_1 \ne \varOmega_2$. The gas pressure at the surface of the core varies quadratically with the cylindrical radius, $R$ (type-V solutions), and $\alpha$ is a free parameter.

In the next section, we examine the mathematical conditions required for the existence of OP, PO, and PP configurations. Throughout this work, we assume $\alpha>1$ (i.e. the core is denser than its envelope), as this is the case in most problems. However, the case $\alpha<1$ is treated in Appendix \ref{app:am}. The results of the two cases are largely the same; only the conditions of positivity of the $\varOmega_i^2$'s differ (see below).

\section{The conditions of positivity of the $\varOmega_i^2$'s}
\label{sec:positivity}

The main constraint is that the rotation rates for the two layers must both be real numbers. From Eqs.~\eqref{eq:om22_sol} and \eqref{eq:om12_sol}, requiring that $\varOmega_i^2\ge0$ leads to the following set of inequalities:
\begin{subequations}\label{eq:om2_pos}
       \begin{empheq}[left={\empheqlbrace\,}]{align}
              &\left[2+5(\alpha-1) q^3\right] \varepsilon_2^2 \geq 3(\alpha-1) q^5\varepsilon_1^2, \label{eq:om22_pos}\\
              &[2+(\alpha-1)(5q^3-3)] \varepsilon_2^2 \geq (\alpha-1)\left[3q^5-2\alpha-3\right]\varepsilon_1^2, \label{eq:om12_pos}
       \end{empheq}
\end{subequations}
which, in standard conditions ($\alpha>1$), becomes
\begin{subequations}\label{eq:om2_pos_ap}
       \begin{empheq}[left={\empheqlbrace\,}]{align}
              &\varepsilon_2^2 \geq \frac{3(\alpha-1) q^5}{2+5(\alpha-1) q^3}\varepsilon_1^2, \label{eq:om22_pos_ap}\\
              &\varepsilon_1^2 \geq -\frac{2+(\alpha-1) (5q^3-3)}{(\alpha-1)\left(3+2\alpha-3q^5\right)}\varepsilon_2^2, \label{eq:om12_pos_ap}
       \end{empheq}
\end{subequations}
as $3q^5-2\alpha-3<0$. We see from Eq.~\eqref{eq:om22_pos_ap} that an oblate core ($\varepsilon_1^2>0$) is only compatible with an oblate envelope ($\varepsilon_2^2>0$). OP configurations are thus ruled out. So we only consider solutions with prolate cores in the following. With the condition $\varepsilon_1^2\leq0$, Eq.~\eqref{eq:om12_pos_ap} leads to 
\begin{equation}
       \left[2 + (\alpha-1)(5q^3-3)\right]\varepsilon_2^2 \geq 0.
\end{equation}
There are therefore two remaining cases, namely the PO and PP configurations.

\subsection{Prolate envelope solution ruled out} With $\varepsilon_2^2\leq 0$, the above inequality reads $2 + (\alpha-1)(5q^3-3)\leq 0$, and so  Eq.~\eqref{eq:om12_pos} becomes\begin{equation}
       \frac{3(\alpha-1) q^5}{2+5(\alpha-1) q^3}\varepsilon_1^2 \leq \varepsilon_2^2 \leq \frac{(\alpha-1)[3q^5-2(\alpha-1)-5]}{2+(\alpha-1)(5q^3-3)}\varepsilon_1^2,
\end{equation}
where we take Eq.~\eqref{eq:om22_pos} into account. As $\varepsilon_1^2\leq0$ is assumed, we find
\begin{equation}
       \frac{3q^5}{2+5(\alpha-1) q^3} \geq \frac{3q^5-2(\alpha-1)-5}{2+(\alpha-1)(5q^3-3)},
\end{equation}
which also reads
\begin{equation}\label{eq:propro_nogo}
       4(\alpha-1) + 10(\alpha-1)^2q^3 + 10 + 25(\alpha-1) q^3 \leq 9(\alpha-1) q^5.
\end{equation}
As $\alpha-1>0$ and given $0<q<1$, we have
\begin{equation}
  {25(\alpha-1) q^3 \geq 9(\alpha-1) q^5},
\end{equation}
which implies that Eq.~\eqref{eq:propro_nogo} cannot be verified. We conclude that a prolate core surrounded by a prolate envelope cannot be at equilibrium. Thus, PP configurations are ruled out as well. We conclude that in the regime of slow rotations, a body made of two rotating, homogeneous layers bound by spheroidal surfaces cannot represent an equilibrium configuration if the envelope is prolate, whatever the shape of the core.

\subsection{Oblate envelope solution permitted} 

If we now impose $\varepsilon_2^2\geq 0$, we see that Eq.~\eqref{eq:om22_pos} is automatically verified. Furthermore, we have $2 + (\alpha-1) (5q^3-3) \geq 0$ in this case, which implies
\begin{equation}\label{eq:qmin}
       q \geq \left[\frac{3 (\alpha-1)-2}{5(\alpha-1)}\right]^{1/3} \equiv q_{\rm lim}
\end{equation}
and
\begin{equation}\label{eq:e2minap1}
       \varepsilon_2^2 \geq \frac{(\alpha-1)\left[3q^5-2(\alpha-1)-5\right]}{2+(\alpha-1)(5q^3-3)}\varepsilon_1^2 \equiv \varepsilon_{2,\lim}^2.
\end{equation}
We see that these two conditions can be simultaneously satisfied in the domain of interest. There is, however, a lower limit for the relative size of the core. We find $q_{\rm lim}=0$ for $\alpha=5/3$, and $q_{\rm lim}=(3/5)^{1/3} \approx 0.843$ for $\alpha \rightarrow \infty$. There is also a lower limit for the eccentricity of the envelope's surface, that is, $\varepsilon_{2,\lim}^2$. Thus, we can state that in the regime of slow rotations, a body made of two rotating layers bound by spheroidal surfaces can be in hydrostatic equilibrium if an oblate envelope surrounds a prolate core.

\subsection{Note on the global rotations}\label{ssec:global_po}

As shown above, the selection of physically relevant configurations is based on the sign of the $\varOmega_i$'s, not on their relative magnitude. As outlined in \citet{h2022a} in relation to OO configurations only, states of global rotation, that is, $\varOmega_1=\varOmega_2$, can occur either for $\alpha=1$, which corresponds to a single-layer object, or for a special mass-density jump, $\alpha_{\rm C}$, namely
\begin{equation}\label{eq:alphac}
       \alpha \equiv \alpha_{\rm C}= 1 + \frac{5(\varepsilon_1^2-\varepsilon_2^2)}{2\varepsilon_1^2+3q^5\varepsilon_1^2-5q^3\varepsilon_2^2}.
\end{equation}
This expression is obtained by equalizing $\tilde\varOmega_1^2$ with the $\tilde\varOmega_2^2$ in Eq.~\eqref{eq:om2_sol} and assuming $\alpha\neq1$. In this case, the rotation rate of the whole structure is given by
\begin{equation}\label{eq:typec}
       \tilde\varOmega^2 = \frac43\frac{\varepsilon_1^2-\varepsilon_2^2}{2\varepsilon_1^2+3q^5\varepsilon_1^2-5q^3\varepsilon_2^2}\varepsilon_1^2 + \frac23\varepsilon_1^2 - \frac25\varepsilon_2^2 \ge 0.
\end{equation}As proved above, only OO and PO configurations are permitted. So, assuming $\alpha\neq 1$, $\varepsilon_1^2\leq0$, and $\varepsilon_2^2\geq 0$, Eq.~\eqref{eq:typec} implies
\begin{equation}
       (20+15q^3)\varepsilon_1^4 - (16+25q^3+9q^5)\varepsilon_1^2\varepsilon_2^2 + 15q^3\varepsilon_2^4 \leq 0.
\end{equation}The left-hand side of this inequality is a second-degree polynomial in $\varepsilon_1^2$. The two roots are 
\begin{align}
       \varepsilon_{1\pm}^2 = &\frac{16+25q^3+9q^5}{10(4+3q^3)}\varepsilon_2^2 \left[1\pm \sqrt{1-\frac{300q^3(4+3q^3)}{(16+25q^3+9q^5)^2}}\right],
\end{align}
and it is easy to see that both roots are positive. As the prefactor of $\varepsilon_1^4$ in the second-degree polynomial is positive, Eq.~\eqref{eq:typec} is verified only between the two roots $\varepsilon_{1\pm}$ (i.e. for $\varepsilon_1^2\geq 0$). This contradicts the starting hypothesis, which means that in the limit of slow rotations, there are no equilibria with a prolate core and an oblate envelope spinning at the same rate. In other words, under our assumptions, the presence of a rotational discontinuity is mandatory for the existence of prolate cores, that is, $\varOmega_1 \ne \varOmega_2$. The only exception is the case $\alpha=1$, where the distinction between the prolate core and the oblate envelope is impossible.

\section{Results}

\subsection{Rotation rates}

As we focused on slow rotations (and small deviations from sphericity), the  $\tilde\varOmega_i$'s are necessarily small values. In contrast, the rotational discontinuity at the interface, defined as
\begin{equation}
       \varDelta = \frac{\tilde\varOmega_1}{\tilde\varOmega_2},
\end{equation}
can take any positive value. From Eqs. \eqref{eq:om22_sol} and \eqref{eq:om12_sol}, we have
\begin{equation}\label{eq:om12_over_om22}
       \varDelta^2 = \frac{1}{\alpha}\left[1 - (\alpha-1)\frac{3\varepsilon_2^2-2(\alpha-1)\varepsilon_1^2-5\varepsilon_1^2}{2\varepsilon_2^2-(\alpha-1)q^3(3q^2\varepsilon_1^2-5\varepsilon_2^2)}\right],
\end{equation}
which is a complicated function of four parameters. However, for any PO configuration, the fraction in Eq.~\eqref{eq:om12_over_om22} is negative, which implies that the term inside the brackets is less than unity. As a consequence, the oblate envelope always rotates faster than the prolate core. 

\begin{figure}[ht]
       \centering
       \includegraphics[width=9.cm]{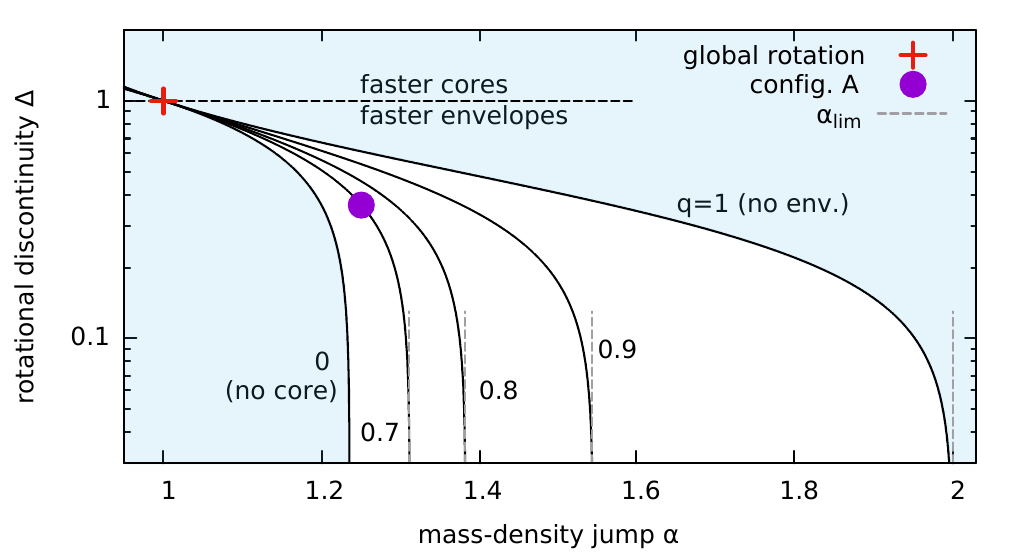}
       
       \includegraphics[width=9.cm]{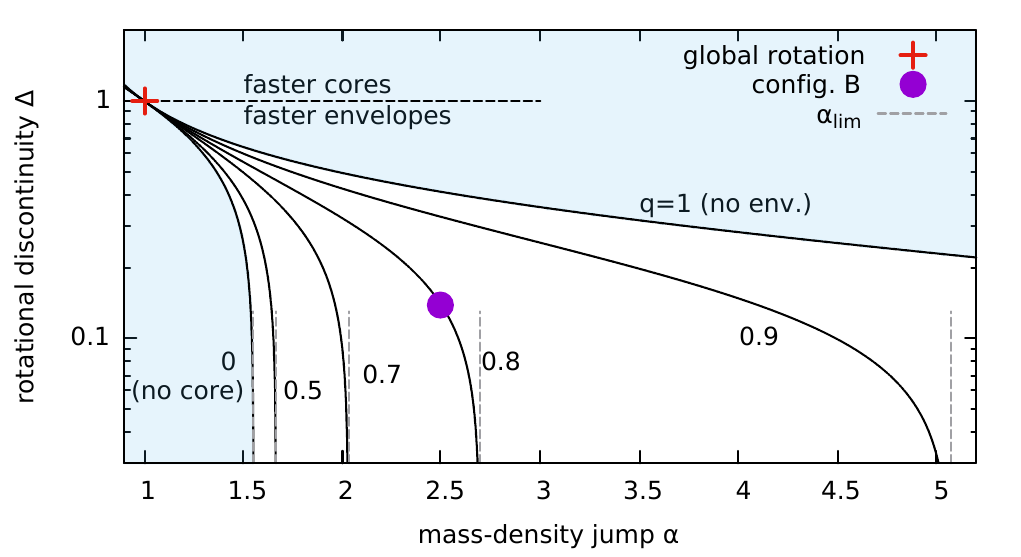}
       \caption{Rotational discontinuity, $\varDelta,$ versus $\alpha$ defined by Eq. \eqref{eq:om12_over_om22} for $\varepsilon_2^2=0.01$ (top) and $\varepsilon_2^2=0.1$ (bottom). In both cases, $\varepsilon_1^2=-0.01$ corresponds to a prolate core surrounded by an oblate envelope. See Tables \ref{tab:configA} and \ref{tab:configB} for configurations A and B.}
       \label{fig:wversusalpha001.pdf}
\end{figure}

Figure \ref{fig:wversusalpha001.pdf} displays $\varDelta$ versus $\alpha$ for two pairs ($\varepsilon_1^2{,}\varepsilon_2^2$). We see that all curves converge to $\varDelta=1$ for $\alpha=1$ because this case corresponds to the single-layer Maclaurin spheroid. Roughly speaking,  larger mass-density jumps are permitted as the oblateness of the envelope increases. Next, we observe that $\varOmega_1 \rightarrow 0$ for some special values of the parameters. This is expected from the equation set. We see from Eq.~\eqref{eq:om2_sol} that the condition $\tilde{\varOmega}_{1}^2=0$ can be obtained for a special value of mass-density contrast $\alpha_{\rm lim}-1 >0$, which is the root of a second-degree polynomial, namely
\begin{equation}
    \varepsilon_{1}^2(\alpha_{\rm lim}{-}1)^2+\frac{5\varepsilon_{1}^2-3\varepsilon_{2}^2-q^3(3q^2\varepsilon_{1}^2{-}5\varepsilon_{2}^2)}{2}(\alpha_{\rm lim}{-}1)+\varepsilon_{2}^2=0.
\end{equation}

It therefore follows that a prolate core can be at rest while the envelope is rotating. In a similar way, the positivity condition $\tilde{\varOmega}_{1}^2 \ge 0$ corresponds to a minimum value for the core's eccentricity, namely
\begin{equation}
\varepsilon_1^2 \ge -\frac{[(\alpha-1)(5q^3-3)+2]\varepsilon_{2}^2}{(\alpha-1)[(5-3q^5) + 2(\alpha-1)]} \equiv \varepsilon^2_{1, \rm min}.
\end{equation}

\subsection{Two examples}

Two-layer configurations hosting prolate cores are easily generated as soon as Eqs.~\eqref{eq:qmin} and \eqref{eq:e2minap1} are simultaneously satisfied. There are four input parameters, namely $q$, $\varepsilon_1$, $\varepsilon_2$, and $\alpha$. We had the opportunity to compare the analytical solution with the equilibrium computed from the {\tt DROP} code, which numerically solves  the axisymmetric problem using the self-consistent field (SCF) method \citep{hachisu86}. This code generates the mass-density and pressure profiles, the equilibrium surfaces, and the interfaces, which generally deviate slightly from spheroids. The performance and reliability of the simulations have already been widely demonstrated \citep{hh17,hhn18}, in particular the multi-layer version \citep{bh21}. To illustrate the calculations reported in this section and the previous one, we considered two examples with large cores. The input and output data are listed in Table \ref{tab:configA} for $\alpha=1.25$, $q=0.7$, and $\varepsilon_2^2=0.01$, and in Table \ref{tab:configB} for $\alpha=2.5$, $q=0.8$, and $\varepsilon_2^2=0.1$. Regarding the SCF code, the numerical experiments were performed with $257$ nodes per direction on a cylindrical $(R,Z)$ grid. To quantify the accuracy of the solutions, we used the relative virial parameter $|{\rm VP}/W|$ \citep[see e.g.][]{ka16,bh21}, where ${{\rm VP} = W+2T+U}$, $W$, $T$, and $U$ being the gravitational, kinetic, and internal energies, respectively. At equilibrium, we should have ${\rm VP} = 0$ \citep[see e.g.][]{cox_1968}. However, due to the numerical treatment ({\tt DROP} is second-order accurate in the grid spacing) and the approximation of NSFoE, ${\rm VP}$ cannot be exactly null in both approaches. The absolute value of this parameter with respect to the greatest energy (i.e. $|W|$) is expected to be of the order of magnitude of the error made with the method.

\begin{table}[ht]
       \centering
       \caption{Example of a configuration and the main output data obtained with a prolate core and an oblate envelope.}
       \begin{tabular}{lrrr}
              \hline\hline
              Config. A &{\tt DROP} & full NSFoE & first-order \\ 
              $\varepsilon_1^2$ & $-0.00979$ & $\leftarrow -0.01$  & $\leftarrow -0.01$ \\  
              $\varepsilon_2^2$ & $\leftarrow 0.01$ & $\leftarrow 0.01$ & $\leftarrow 0.01$ \\ 
              $q$ & $0.70007$ & $\leftarrow 0.70$ & $\leftarrow 0.70$ \\  
              $\alpha$ & $\leftarrow 1.25$ & $\leftarrow 1.25$ & $\leftarrow 1.25$ \\
              $-c$ & $1.685\times10^{-2}$ & $1.700\times10^{-2}$ & $1.700\times10^{-2}$ \\
              $\varepsilon_{2,\lim}^2$ & & & $7.439\times10^{-3}$ \\
              $q_{\rm lim}$ & & & $-1.00000$ \\
              $b_1/a_2$ & $\leftarrow {\rm NSFoE}$ & $0.70349$ & $0.70349$ \\ 
              $\tilde\varOmega_1^2$ & $4.585\times10^{-4}$ & $4.596\times10^{-4}$ & $4.585\times10^{-4}$ \\ 
              $\tilde\varOmega_2^2$ & $3.411\times10^{-3}$ & $3.420\times10^{-3}$ & $3.406\times10^{-3}$ \\ 
              $\varDelta$ & $\leftarrow  {\rm NSFoE}$ & $3.666\times10^{-1}$ & $3.669\times10^{-1}$ \\
              $M/(\rho_2a_2^3)$ & $4.52913$ & $4.52877$ & $4.52883$ \\
              $p_{\rm c}/(\uppi G\rho_2^2a_2^2)$ & $8.963\times10^{-1}$ & $8.963\times10^{-1}$ & $8.962\times10^{-1}$ \\
              $|{\rm VP}/W|$ & $7\times10^{-7}$ & $4\times10^{-6}$ & $3\times10^{-3}$ \\
              \hline
       \end{tabular}
       \tablefoot{
              $p_{\rm c}$ is the central pressure and $M$ is the mass; `$\leftarrow$'~indicates input data. The second column contains the results from the numerical solution obtained with the {\tt DROP} code \citep{bh21}. The third column contains the results from the full NSFoE theory \citep{h2022a} and the fourth the results obtained with the first-order approximation of $\varepsilon^2$ used in this work.
       }
       \label{tab:configA}
\end{table}

\begin{table}[ht]
       \centering
       \caption{Same as Table 1  but for configuration B.}
       \begin{tabular}{lrrr}
              \hline\hline   
              Config. B &{\tt DROP} &full NSFoE & this work\\   
              $\varepsilon_1^2$ & $-0.01025$ & $\leftarrow -0.01$ & $\leftarrow -0.01$\\
              $\varepsilon_2^2$ & $\leftarrow 0.1$ & $\leftarrow 0.1$ & $\leftarrow 0.1$\\ 
              $q$ & $0.79990$ & $\leftarrow 0.8$ & $\leftarrow 0.8$\\ 
              $\alpha$ & $\leftarrow 2.5$ & $\leftarrow 2.5$ & $\leftarrow 2.5$\\ 
              $-c$ & $1.082\times10^{-2}$ & $1.080\times10^{-2}$ & $1.080\times10^{-2}$ \\
              $\varepsilon_{2,\lim}^2$ & & & $7.855\times10^{-2}$\\ 
              $q_{\rm lim}$ & & & $6.934\times10^{-1}$\\ 
              $b_1/a_2$ & $\leftarrow{\rm NSFoE}$ & $8.040\times10^{-1}$ & $8.040\times10^{-1}$\\ 
              $\tilde\varOmega_1^2$ & $1.925\times10^{-3}$ & $1.931\times10^{-3}$ & $1.533\times10^{-3}$ \\ 
              $\tilde\varOmega_2^2$ & $8.469\times10^{-2}$ & $8.491\times10^{-2}$ & $7.983\times10^{-2}$ \\ 
              $\varDelta$ & $\leftarrow {\rm NSFoE}$ & $1.508\times10^{-1}$ & $1.385\times10^{-1}$\\ 
              $M/(\rho_2a_2^3)$ & $7.20079$ & $7.20687$ & $7.21242$\\ 
              $p_{\rm c}/(\uppi G\rho_2^2a_2^2)$ & $3.10451$ & $3.10511$ & $3.16290$\\
              $|{\rm VP}/W|$ & $2\times10^{-5}$ & $7\times10^{-4}$ & $6\times10^{-2}$ \\
              \hline
       \end{tabular}
       \label{tab:configB}
\end{table}

As we can see by comparing columns 2 and 4 of Tables \ref{tab:configA} and \ref{tab:configB}, the prolate cores predicted by the calculus are also observed in the numerical experiments, and the axis ratio (which is an output for {\tt DROP}) coincides with the value of the analytical solution within the numerical precision. The virial parameters are small enough that we can assert that the prolateness of the core is not an artefact of the numerical treatment. For configuration A, the values of the rotation rates are also in good agreement, and this is true for the other global quantities as well. For configuration B, the accordance is a little less satisfactory. This is due to the fact that Eqs.~\eqref{eq:om22_sol} and \eqref{eq:om12_sol} are first-order accurate in the eccentricities, while $\varepsilon_2^2=0.1$ is marginally acceptable regarding this expansion. The third columns of Tables \ref{tab:configA} and \ref{tab:configB} show the results obtained via the analytical approach, but with the full expressions of the rotation rates. We see that the agreement between the full NSFoE theory and the numerical SCF method is remarkable, and this confirms the existence of PO configurations.

\section{Discussion and perspectives}

We examined the conditions of the hydrostatic equilibrium of a heterogeneous body made of two rigidly rotating, homogeneous layers bound by a spheroidal surface. Given our hypothesis, and in particular the limit of slow rotations, we find a new family of solutions that involves a prolate core. We have shown that:
\begin{itemize}
\item The existence of a prolate core requires an asynchronous motion between the two layers.
\item The envelope is necessarily oblate.
\item The envelope spins faster than the core.
\item The prolate core can be at rest (for some special parameters).
\item Global rotation is only compatible with an oblate core surrounded by an oblate envelope.
\end{itemize}

\noindent These results are fully confirmed by numerical experiments based on the SCF method.

\subsection{Prolate cores in compressible systems}

A natural question concerns the extension of the results reported above to the case where the mass densities are no longer uniform in each layer. This is easily checked by performing numerical simulations with appropriate equations of state. For this purpose, we used the {\tt DROP} code again, and a polytropic equation of state where the gas pressure, $p,$ is a power law of the mass density, $\rho$, namely $p \propto \rho^{1+1/n}$, and $n$ is the polytropic index. Table \ref{tab:configCD} presents two new configurations, namely:
\begin{itemize}
\item Configuration C, which resembles configuration A but with $(n_1,n_2)=(3,1.5)$. These parameters are appropriate for a radiative pressure-dominated core surrounded by a convective envelope. The mass-density jump is small, and the envelope spins faster than the core by a factor of about $2$.
\item Configuration D, which resembles configuration B but with $(n_1,n_2)=(0.3,0.5)$. This case corresponds to weakly compressible layers. The mass-density jump is twice as large as for configuration C, and the envelope spins faster than the core by a factor of about $3$.
\end{itemize}

\begin{figure}
       \includegraphics[trim={7cm 13.5cm 8cm 0},clip,width=\linewidth]{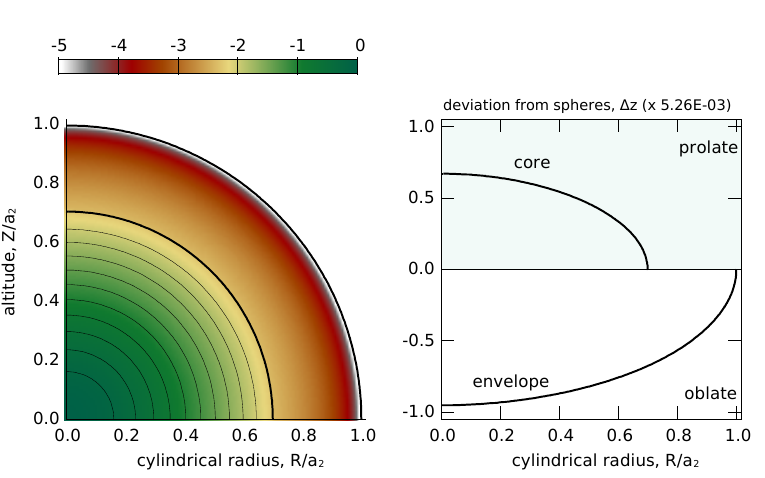}
       \caption{Equilibrium solution computed with the {\tt DROP} code for configuration C: the normalized mass-density, $\rho/\rho_{\rm c}$, map in colour code (left) and the deviation from sphericity (right). See also Table \ref{tab:configCD}.}
       \label{fig:configC}
\end{figure}

\begin{figure}
\includegraphics[trim={7cm 13.5cm 8cm 0},clip,width=\linewidth]{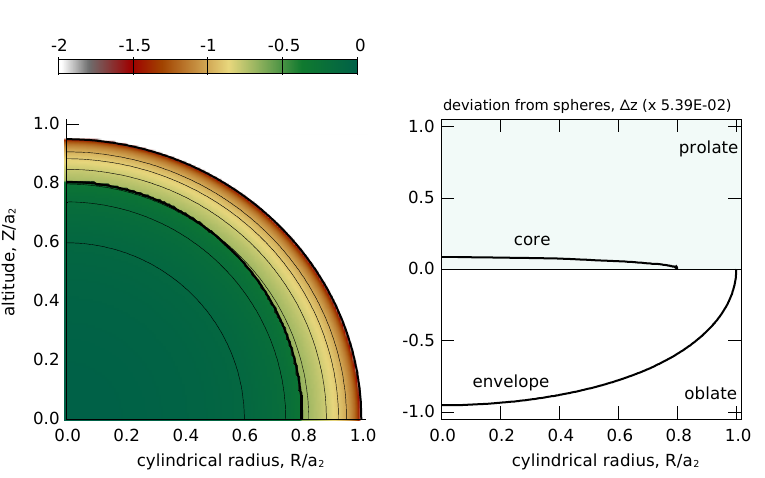}
\caption{Same as Fig.~\ref{fig:configC} but for configuration D.}
\label{fig:configD}
\end{figure}

\begin{table}[ht]
       \centering
       \caption{Same as Table \ref{tab:configA} but for fully heterogeneous layers, computed with the {\tt DROP} code; the solutions obtained are shown in Figs.~\ref{fig:configC} and \ref{fig:configD}.}
       \begin{tabular}{lrr}
              \hline\hline
              &Config. C & Config. D \\
              $\varepsilon_1^2$ & $-0.00996$ & $-0.01039$ \\
              $\varepsilon_2^2$ & $\leftarrow 0.01$ &  $\leftarrow 0.1$\\ 
              $n_1$ & $\leftarrow 3.0$ & $\leftarrow 0.3$\\
              $n_2$ & $\leftarrow 1.5$ & $\leftarrow 0.5$ \\
              $q$ & $0.70001$ & $0.79985$ \\
              $b_1/a_2$ & $\leftarrow0.70349$ & $\leftarrow0.80399$\\ 
              $\alpha$ & $\leftarrow 1.25$ & $\leftarrow 2.5$ \\
              $\varOmega_1^2/(2\uppi G\rho_{\rm c})$ & $3.420\times10^{-5}$ & $2.667\times10^{-3}$\\ 
              $\varOmega_2^2/(2\uppi G\rho_{\rm c})$ & $1.710\times10^{-4}$ & $2.614\times10^{-2}$\\ 
              $\varDelta$ & $\leftarrow 0.44721$ & $\leftarrow 0.31937$ \\ 
              $M/(\rho_{\rm c} a_2^3)$ & $0.11012$ & $1.86401$\\
              $p_{\rm c}/(G\rho_{\rm c}^2a_2^2)$ & $8.352\times10^{-2}$ & $1.02538$\\
              $|{\rm VP}/W|$ & $7\times10^{-5}$ & $6\times10^{-4}$ \\
              \hline
       \end{tabular}
       \tablefoot{$\rho_{\rm c}$ is the central mass density.}
       \label{tab:configCD}
\end{table}

Figures \ref{fig:configC} and \ref{fig:configD} show the mass density and the deviations of the surface bounding layers from spheres. The right panels shows the deviation $\Delta Z_i(R)$ of the true surfaces from spheres, namely 
\begin{equation}
       \Delta Z_i = Z_i(R) - \sqrt{a_i^2-R^2}, \qquad i\in\{1,2\},
\end{equation}
where $Z_i$ is the equation of the boundary of layer $i$, and $R$ is the distance to the rotation axis. Negative (positive) values correspond to oblate (prolate) surfaces. For these two configurations, we observe that the core-envelope interface is indeed prolate in shape (while isobars remain oblate).

\subsection{Possible cancellation of the gravitational moments}

A prolate core embedded in an oblate envelope leaves a mark in the gravitational potential, as it mitigates the effect of an oblate envelope. It can be shown that one of the gravitational moments, $J_{2n}$, can vanish (when $n$ is an odd number). The first coefficient, $J_2$, is of special interest because it is the main deviation from sphericity and therefore of interest for dynamical studies. For spheroidal heteroeoids, the $J_{2n}$'s are analytical \citep{heiskanen1967,cmm19,bh23}. We find that $J_2=0$ when 
\begin{equation}
       \varepsilon_2^2\sqrt{1-\varepsilon_2^2}+(\alpha-1)q^5 \varepsilon_1^2\sqrt{1-\varepsilon_1^2}=0,
\end{equation}
which is a formally cubic equation in $\varepsilon_1^2$. In the actual conditions of slow rotation, this implies
  \begin{equation}
    \varepsilon_1^2 \approx -\frac{\varepsilon_2^2}{(\alpha-1)q^5} \leq 0.
  \end{equation}
This concerns a wide range of equilibria, from large cores and small mass-density jumps to small cores with very high $\alpha$'s, typically.

\subsection{Extension to fast rotators}

An interesting point concerns the extension of the results to the case of fast rotations, which might reveal the existence of very elongated prolate cores. By construction, the NSFoE theory is valid for highly flattened systems, provided the confocal parameter (see Sect. 3) is small enough. It then fully applies to structures elongated along the axis of rotation. We surveyed the $(\varepsilon_1^2,\varepsilon_2^2)$ plane, with $-1\leq\varepsilon_i^2\leq1$, for several values of $q$ and $\alpha$. The main conclusions of this additional exploration are:
\begin{itemize}
       \item Again, there is no solution with a prolate envelope, regardless of the shape and size of the core.
       \item The permitted range of prolateness for the core gets wider as $\alpha$ gets closer to $1$ and/or $q$ decreases.
       \item The permitted range of prolateness for the core gets wider as the envelope gets more flattened.
\end{itemize}

\begin{table}[ht]
       \centering
       \caption{Same as Table \ref{tab:configA} but for a fast-rotating envelope.}
       \begin{tabular}{lrr}
              \hline\hline
              Config. E &{\tt DROP} & full NSFoE \\ 
              $\varepsilon_1^2$ & $-0.26394$ & $\leftarrow -0.26394$ \\  
              $\varepsilon_2^2$ & $\leftarrow 0.37589$ & $\leftarrow 0.37589$ \\ 
              $q$ & $0.66709$ & $\leftarrow 0.66709$ \\  
              $\alpha$ & $\leftarrow 1.25$ & $\leftarrow 1.25$ \\
              $-c$ & $4.933\times10^{-1}$ & $4.933\times10^{-1}$ \\
              $b_1/a_2$ & $\leftarrow 0.75$ & $0.74998$ \\ 
              $\tilde\varOmega_1^2$ & $2.811\times10^{-2}$ & $2.934\times10^{-2}$ \\ 
              $\tilde\varOmega_2^2$ & $1.406\times10^{-1}$ & $1.419\times10^{-1}$ \\ 
              $\varDelta$ & $\leftarrow  0.44721$ & $4.547\times10^{-1}$ \\ 
              $M/(\rho_2a_2^3)$ & $3.62718$ & $3.65865$ \\
              $p_{\rm c}/(\uppi G\rho_2^2a_2^2)$ & $7.232\times10^{-1}$ & $7.236\times10^{-1}$  \\
              $|{\rm VP}/W|$ & $4\times10^{-5}$ & $8\times10^{-3}$ \\
              \hline
       \end{tabular}
       \label{tab:configE}
\end{table}

An example of a fast-rotating envelope surrounding a prolate core, obtained from both the {\tt DROP} code and NSFoE, is given in Table \ref{tab:configE}. We see that the confocal parameter is marginally acceptable in this case ($|c|\sim0.5$), but the analytical solution is still in good agreement with the numerical experiment.

\subsection{Note on the problem of friction between layers}

A major question concerns the time stability of the asynchronously rotating layers, even in the limit of slow rotation. The existence of a permanent difference in the velocities at the core-envelope interface is a source of friction, local heating, and energy dissipation, at the expense of the relative motion. If there is no mechanism able to maintain the difference in the rotation rates, then the shear is expected to brake the envelope and accelerate the core through a boundary layer until synchronicity, that is, $\varDelta \rightarrow 1$. The problem of friction between asynchronously rotating, spherical layers was studied in detail by \citet{wavre25} for viscous forces at the interface of the form $\kappa |\varOmega_1-\varOmega_2|^mR^m$, where $\kappa$ is a friction constant and $m$ the friction index, both positive quantities. For a Stokes-like viscosity, that is, $m=1$, Wavre showed that the relative rotation rate $|\varOmega_1-\varOmega_2|$ between the spheres undergoes an exponential decrease with a characteristic time, $\tau,$ given by
\begin{equation}\label{eq:tau_wavre}
       \frac{1}{\tau} = \frac{8}{3}\uppi a_1^4 \kappa \frac{I_1+I_2}{I_1I_2},
\end{equation}
where $I_1$ and $I_2$ are the moments of inertia of layers 1 and 2 with respect to the rotation axis, respectively. As the deviation of the core-envelope interface from a sphere is assumed to be weak, Eq.~\eqref{eq:tau_wavre} corresponds to the zeroth-order term in $\varepsilon_1^2$, which is enough to obtain an approximate order of magnitude of $\tau$. The main difficulty of this estimation of the characteristic time is the value of $\kappa$. For a fluid core embedded in a precessing mantle, we have $\kappa \propto I_1\sqrt{\nu \varOmega_2}/a_1^5$ \citep{bu68}. However, this expression is probably not relevant for non-precessing, asynchronously rotating layers. Moreover, the viscosity is often unknown in astrophysical contexts. Even for the outer core of the Earth, the uncertainty on the kinematic viscosity, $\nu$, covers 13 orders of magnitude \citep{la91}, the best estimate being $\nu \sim 10^{-6}~\mathrm{m^2\cdot s^{-1}}$ \citep{ga72,po88}. To evaluate the order of magnitude of $\tau$, we assumed that the expression of \citet{bu68} is valid in our case and that $\varOmega_2$ is the initial value of the rotation rate of the envelope, namely
\begin{equation}
      \tau \approx \frac{1}{2.62}\frac{a_1}{(2\uppi G \rho_2 \nu^2)^{1/4}} \frac{1}{\sqrt{\tilde\varOmega_{2}}}\frac{1-q^5}{1+(\alpha-1)q^5}.
\end{equation}
For the parameters of configuration A in Table \ref{tab:configA}, taking ${\nu = 10^{-6}~\mathrm{m^2\cdot s^{-1}}}$ and assuming a planetary core with $a_1=1000~\mathrm{km}$ and a mass density of the envelope $\rho_2=4000~\mathrm{kg\cdot m^{-3}}$ (typically silicates), we have $\tau \sim 10^3~\mathrm{yr}$. Such a small value is expected, as this situation corresponds roughly to a slightly differentiated rocky planet; the friction between the components is thus strong. The situation is probably more favourable in the interstellar medium, where the components are mostly gaseous and so the friction is weaker. Unfortunately, the viscosity is even less constrained in such contexts.

\begin{acknowledgements}
       We are grateful to the referee, F. Chambat, for his suggestions and remarks to improve the paper, in particular for clarifications on the classical works of Clairaut, Poincar\'e and Hamy, and for pointing out the monograph of D'Alembert. We also thank P. Auclair-Desrotour for interesting references on the problem of friction at planetary core-mantle boundaries.
\end{acknowledgements}

\bibliographystyle{aa}

\bibliography{problate}

\begin{thebibliography}{30}
\expandafter\ifx\csname natexlab\endcsname\relax\def\natexlab#1{#1}\fi

\bibitem[{{Basillais} \& {Hur{\'e}}(2021)}]{bh21}
{Basillais}, B. \& {Hur{\'e}}, J.-M. 2021, \mnras, 506, 3773

\bibitem[{{Basillais} \& {Hur{\'e}}(2023)}]{bh23}
{Basillais}, B. \& {Hur{\'e}}, J.-M. 2023, \mnras, 520, 1504

\bibitem[{{Busse}(1968)}]{bu68}
{Busse}, F.~H. 1968, Journal of Fluid Mechanics, 33, 739

\bibitem[{{Cai} \& {Taam}(2010)}]{ct10}
{Cai}, M.~J. \& {Taam}, R.~E. 2010, \apjl, 709, L79

\bibitem[{{Cisneros-Parra} {et~al.}(2019){Cisneros-Parra}, {Martinez-Herrera},
  \& {Montalvo-Castro}}]{cmm19}
{Cisneros-Parra}, J.~U., {Martinez-Herrera}, F.~J., \& {Montalvo-Castro}, J.~D.
  2019, \apjs, 241, 8

\bibitem[{Clairaut(1743)}]{clairaut43}
Clairaut, A.~C. 1743, Th{\'e}orie de la figure de la Terre tir{\'e}e des
  principes de l'hydrostatique (Paris: David Fils)

\bibitem[{Cox \& Giuli(1968)}]{cox_1968}
Cox, J.~P. \& Giuli, R.~T. 1968, Principles of stellar structure {Volume} {I} :
  {Physical} principles, 1st edn. (New York: Gordon and Breach)

\bibitem[{D'Alembert(1747)}]{dalembert47}
D'Alembert, J. 1747, R\'eflexions sur la cause g\'en\'erale des vents (Paris:
  David l'a\^in\'e)

\bibitem[{{Fiege} \& {Pudritz}(2000)}]{fp00}
{Fiege}, J.~D. \& {Pudritz}, R.~E. 2000, \apj, 534, 291

\bibitem[{{Fujisawa} \& {Eriguchi}(2014)}]{fe14}
{Fujisawa}, K. \& {Eriguchi}, Y. 2014, \mnras, 438, L61

\bibitem[{{Gans}(1972)}]{ga72}
{Gans}, R.~F. 1972, \jgr, 77, 360

\bibitem[{{Hachisu}(1986)}]{hachisu86}
{Hachisu}, I. 1986, \apjs, 61, 479

\bibitem[{{Hamy}(1889)}]{hamy89}
{Hamy}, M. 1889, Annales de l'Observatoire de Paris, 19, F.1

\bibitem[{Hamy(1890)}]{hamy90}
Hamy, M. 1890, Journal de math\'ematiques pures et appliqu\'ees 4e s\'erie, 6,
  69

\bibitem[{Heiskanen \& Moritz(1967)}]{heiskanen1967}
Heiskanen, W. \& Moritz, H. 1967, Physical Geodesy (San Fransisco: W.H.
  Freeman)

\bibitem[{{Hur{\'e}}(2022)}]{h2022a}
{Hur{\'e}}, J.-M. 2022, \mnras, 512, 4031

\bibitem[{{Hur{\'e}} \& {Hersant}(2017)}]{hh17}
{Hur{\'e}}, J.-M. \& {Hersant}, F. 2017, \mnras, 464, 4761

\bibitem[{{Hur{\'e}} {et~al.}(2018){Hur{\'e}}, {Hersant}, \& {Nasello}}]{hhn18}
{Hur{\'e}}, J.-M., {Hersant}, F., \& {Nasello}, G. 2018, \mnras, 475, 63

\bibitem[{Hur\'e \& Staelen(2024)}]{hs24}
Hur\'e, J.-M. \& Staelen, C. 2024, Phys. Rev. D, 110, 063017

\bibitem[{{Kadam} {et~al.}(2016){Kadam}, {Motl}, {Frank}, {Clayton}, \&
  {Marcello}}]{ka16}
{Kadam}, K., {Motl}, P.~M., {Frank}, J., {Clayton}, G.~C., \& {Marcello}, D.~C.
  2016, \mnras, 462, 2237

\bibitem[{{Kawamura} {et~al.}(2011){Kawamura}, {Taniguchi}, {Yoshida}, \&
  {Eriguchi}}]{ktye11}
{Kawamura}, T., {Taniguchi}, K., {Yoshida}, S., \& {Eriguchi}, Y. 2011, \mnras,
  416, L75

\bibitem[{{Lander} \& {Jones}(2009)}]{lj09}
{Lander}, S.~K. \& {Jones}, D.~I. 2009, \mnras, 395, 2162

\bibitem[{{Lumb} \& {Aldridge}(1991)}]{la91}
{Lumb}, L.~I. \& {Aldridge}, K.~D. 1991, Journal of geomagnetism and
  geoelectricity, 43, 93

\bibitem[{{Montalvo} {et~al.}(1983){Montalvo}, {Martinez}, \&
  {Cisneros}}]{mmc83}
{Montalvo}, D., {Martinez}, F.~J., \& {Cisneros}, J. 1983, \rmxaa, 5, 293

\bibitem[{Moulton(1916)}]{moulton16}
Moulton, E.~J. 1916, Transactions of the American Mathematical Society, 17, 100

\bibitem[{{Oganesyan} \& {Abramyan}(1972)}]{oa72}
{Oganesyan}, R.~S. \& {Abramyan}, M.~G. 1972, Astrophysics, 8, 352

\bibitem[{{Poirier}(1988)}]{po88}
{Poirier}, J.~P. 1988, Geophysical Journal International, 92, 99

\bibitem[{{Rambaux} {et~al.}(2015){Rambaux}, {Chambat}, \&
  {Castillo-Rogez}}]{rcc15}
{Rambaux}, N., {Chambat}, F., \& {Castillo-Rogez}, J.~C. 2015, \aap, 584, A127

\bibitem[{V\'eronet(1912)}]{veronet12}
V\'eronet, A. 1912, Journal de math{\'e}matiques pures et appliqu{\'e}es 6e
  s\'erie, 8, 331

\bibitem[{Wavre(1925)}]{wavre25}
Wavre, R. 1925, Archive des sciences physiques et naturelles 5me p\'eriode, 7,
  133

\end{thebibliography}

\begin{appendix}

\section{The case $\alpha<1$}\label{app:am}

Considering $\alpha<1$, the system of inequalities \eqref{eq:om2_sol} becomes
\begin{subequations}\label{eq:om2_pos_am}
       \begin{empheq}[left={\empheqlbrace\,}]{align}
              &\varepsilon_1^2\geq -\frac{2-5(1-\alpha)q^3}{3(1-\alpha)q^5}\varepsilon_2^2, \label{eq:om22_pos_am}\\
              &\frac{2-(1-\alpha)[5q^3-3]}{(1-\alpha)\left[5-2(1-\alpha)-3q^5\right]}\varepsilon_2^2 \geq \varepsilon_1^2 \label{eq:om12_pos_am}
       \end{empheq}
\end{subequations}

Combining the two inequalities of the system, we readily obtain
\begin{equation}
       \frac{2-(1-\alpha)[5q^3-3]}{(1-\alpha)\left[5-2(1-\alpha)-3q^5\right]}\varepsilon_2^2 \geq -\frac{2-5(1-\alpha)q^3}{3(1-\alpha)q^5}\varepsilon_2^2
\end{equation}

As before, we now have two cases to consider on the sign of $\varepsilon_2^2$.

\subsection{Prolate envelope ruled out}
       
We first assume $\varepsilon_2^2<0$. We thus have
\begin{equation}
       -\frac{2-5(1-\alpha)q^3}{3(1-\alpha)q^5}\geq \frac{2-(1-\alpha)[5q^3-3]}{(1-\alpha)\left[5-2(1-\alpha)-3q^5\right]}
\end{equation}
which leads to
\begin{equation}
       10 - [4+25q^3-9q^5](1-\alpha) + 10(1-\alpha)^2 \leq 0.
\end{equation}
As $4+25q^3-9q^5<20$, the discriminant of the second-degree polynomial in $1-\alpha$ in the left hand side is negative, except for $q=1$. In this case, however, we see that the inequality becomes $\alpha^2\leq0$, which is not possible. Thus, in a slightly flattened, heterogeneous system made of two homogeneous components, there exists no equilibrium with a prolate envelope.

\subsection{Prolate core, oblate envelope permitted}
       
In this study, we are interested in prolate subsystems. We then impose $\varepsilon_1^2\leq0$. Thus, Eq. \eqref{eq:om12_pos_am} is automatically verified and Eq. \eqref{eq:om22_pos_am} reads
\begin{equation}\label{eq:e2minam1}
       \varepsilon_2^2 \geq -\frac{3(1-\alpha)q^5}{2-5(1-\alpha)q^3}\varepsilon_1^2,
\end{equation}
where we have necessarily $2-5(1-\alpha)q^3\geq0$, meaning that we have an upper limit on $q$ (whereas we had a lower limit in the $\alpha>1$ case), namely
\begin{equation}\label{eq:qmax}
       q \leq \left(\frac{2}{5(1-\alpha)}\right)^{1/3} \equiv q_{\lim}'.
\end{equation}
Thus, in a slightly flattened, heterogeneous system made of two homogeneous components, there exists equilibria with an oblate envelop and a less dense, prolate core.

\end{appendix}

\label{lastpage}

\end{document}